\documentclass[lineno]{jfm}

\usepackage{graphicx}
\usepackage{newtxtext}
\usepackage{newtxmath}
\usepackage{natbib}
\usepackage[dvipsnames]{xcolor}
\usepackage{fdsymbol}
\usepackage{pifont}
\usepackage{hyperref}
\usepackage{url}
\newcommand{\RomanNumeralCaps}[1]
\linenumbers

\newcommand{\re}{Re_{\tau}}
\newcommand{\ks}{k_{S}^{+}}
\newcommand{\ds}{\delta_{S}}
\newcommand{\ie}{\emph{i.e.},\ }
\newcommand{\egg}{\emph{e.g.},\ }

\definecolor{mgreen}{rgb}{0, 0.6, 0}
\definecolor{myellow}{rgb}{1, 0.8, 0}
\definecolor{mpurple}{rgb}{0.7, 0, 0.7}

\newcommand{\rev}[1]{{\color{black}#1}}
\newcommand{\rew}[1]{{\color{black}#1}}
\newcommand{\rex}[1]{{\color{black}#1}}

\title{Defining the mean turbulent boundary layer thickness based on streamwise velocity skewness}

\author{Mitchell Lozier$^1$
  \corresp{\email{mitchell.lozier@unimelb.edu.au}},
  Rahul Deshpande$^1$, Ahmad Zarei$^1$, Luka Lindi\'c$^1$,
  Wagih Abu Rowin$^1$ \and Ivan Marusic$^{1,2}$}

\affiliation{$^1$Department of Mechanical Engineering, The University of Melbourne, Australia\\
$^2$Faculty of Mechanical Engineering and Naval Architecture, University of Zagreb, Croatia}

\begin{document}
\maketitle

\begin{abstract}
\rev{A new statistical definition for the mean turbulent boundary layer thickness is introduced, based on identification of the wall-normal location where the streamwise velocity skewness changes sign, from negative to positive, in the outermost region of the boundary layer.} 
\rew{Importantly, this definition is independent of arbitrary thresholds, and broadly applicable, including to past single-point measurements. 
Further, this definition is motivated by the phenomenology of streamwise velocity fluctuations near the turbulent/non-turbulent interface, whose local characteristics are shown to be universal for turbulent boundary layers under low freestream turbulence conditions (\ie with or without pressure gradients, surface roughness, etc.) through large-scale experiments, simulations and coherent structure-based modelling. 
The new approach yields a turbulent boundary layer thickness that is consistent with previous definitions, such as those based on Reynolds shear stress or `composite' mean velocity profiles, and which can be used practically \egg to calculate integral thicknesses.} 
Two methods are proposed for estimating the turbulent boundary layer thickness using this definition: one based on simple linear interpolation and the other on fitting a generalised Fourier model to the outer skewness profile. 
The robustness and limitations of these methods are demonstrated through analysis of several published experimental and numerical datasets, which cover a range of canonical and non-canonical turbulent boundary layers. 
These datasets also vary in key characteristics such as wall-normal resolution and measurement noise, particularly in the critical turbulent/non-turbulent interface region. 
\end{abstract}



\section{Introduction}
\label{sec:intro}

Determining the relevant characteristic length scales of turbulent flows is critical for both characterising their state, and describing their development. 
The outer length scale is of particular interest, as it defines the transverse extent of a turbulent flow, and consequently, the maximum size of turbulent motions, or eddies, within. 
Here, we limit our focus to the turbulent boundary layer (TBL), which governs the performance of a range of engineering systems, and where the outer length scale is generally referred to as the TBL thickness, $\delta$. 
Unlike internal flows, where the outer length scale is defined explicitly by geometric constraints (\egg the channel mid-height or pipe radius) the TBL is only semi-constrained, with its wall-normal extent inferred from a pair of `boundaries'. 
The first boundary is a solid wall, which typically has well-defined boundary conditions. 
The second boundary is a complex, freely developing, three-dimensional interface between turbulent eddies within the TBL and the external freestream flow, broadly referred to as the turbulent non-turbulent interface (TNTI). 
Locally, the wall-normal distance between the solid wall and this freely developing interface represents an instantaneous thickness of the TBL. 
While the instantaneous thickness is finite and relatively simple to visualise (\egg see flow visualisations from \citealp{Baxerres_PGBoundaryLayers_2024}), defining an outer length scale instantaneously is not practical due to the stochastic nature of the TBL \citep{daSilva_InterfacialLayers_2014, Reuther_TNTI_2018}. 
In light of this, we seek an average outer length scale which is characteristic of the converged TBL statistics. 
However, rigorously quantifying this characteristic outer length scale in a flow with such a complex and stochastic interfacial boundary has remained persistently difficult, in contrast to pipe or channel flows for instance. 
This has led to the proposal of many statistical approaches/methods to estimate this characteristic outer length scale, $\delta$, some of which are summarised in table~\ref{tab:deltas} for reference and discussed in detail below. 

\begin{table}
  \begin{center}
  \begin{tabular}{ccccc}
    Method Name & Equation & Flow Properties & Threshold? & Terminology \\ \hline
    $99\%$ & \eqref{eq:d99} & $U$ & Yes & $\delta_{99}$ \\
    Composite Profile & \eqref{eq:d125} & $U$ & Yes & $\Delta_{1.25}$ \\
    Diagnostic Plot & \eqref{eq:dD} & $ \overline{u^{2}} \&\ U$ & Yes & $\delta_{D}$ \\
    Reynolds Shear Stress & \eqref{eq:duw} & $\overline{uw}$ & Yes & $\delta_{uw}$ \\
    TNTI & \eqref{eq:dtnti} & $\tilde{k}$ & Yes & $\delta_{\mathrm{TNTI}}$ \\ \hline
  \end{tabular}
  \caption{Summary of common TBL thickness estimation methods.}
  \label{tab:deltas}
  \end{center}
\end{table}

Perhaps the earliest and most prolific method for estimating the TBL thickness is the $99\%$ thickness \citep{Schlichting_BLTheory_1955}, commonly referred to as $\delta_{99}$, where 
\begin{equation}
  U(z=\delta_{99})=0.99U_{\infty}.
  \label{eq:d99}
\end{equation}
Here, $U_{\infty}$ is the freestream velocity, $U$ is the mean streamwise velocity, $z$ is the wall-normal distance from the wall (with $z=0$ being the wall), and $\delta_{99}$ is defined as the wall-normal distance where $U$ reaches $99\%$ of $U_\infty$. 
This method has been used extensively and is relatively simple to implement in both experiments and simulations. 
However, the prescribed threshold of $99\%$ of the freestream velocity is somewhat arbitrary.
Some studies \citep[see][]{Kundu_Fluids_1990} have considered relatively stricter or more lenient thresholds, such as $99.5\%$ or $95\%$ of the freestream velocity, respectively, highlighting the ambiguity in this method. 
In some cases, the presumed asymptotic decay of the mean shear ($\mathrm{d}U/\mathrm{d}z$) has also been considered as an alternative metric to estimate the TBL thickness, but this method suffers from the same ambiguity in determination of an appropriate threshold. 

One solution to this ambiguity is the use of composite mean streamwise velocity profiles, specifically for the wake region of the TBL, which have been used in several studies to estimate a representative mean TBL thickness \citep{Coles_WakeParam_1956, Nickels_InnerScaling_2004, Chauhan_CriteriaZPG_2009}. 
Recently, \citet{Baxerres_PGBoundaryLayers_2024} reported the TBL thickness found using these composite profiles ($\Delta_{1.25}$) to be approximately related to $\delta_{99}$ by a constant:
\begin{equation}
  \Delta_{1.25} = 1.25\delta_{99}.
  \label{eq:d125}
\end{equation}
These composite profiles also typically incorporate the assumed asymptotic behaviour of the canonical mean streamwise velocity profile in their formulation. 
In the case of non-canonical TBLs, however, the asymptotic behaviour may differ from the canonical case near the TBL edge. 
For instance, it has been reported for adverse-pressure gradient (APG) TBLs that the mean shear is not guaranteed to be zero above $\delta$ (\ie $dU/dz(z>\delta)\neq 0$; \citealp{Vinuesa_LengthScale_2016, Griffin_BLThickness_2021}). 
As such, reliance on canonical composite/wake profiles, and specifically the assumptions made about the mean shear and/or the mean streamwise velocity behaviour in the outer region of canonical TBLs, has left an open question about the use of these methods for non-canonical TBLs. 
To that end, a revised definition of the TBL thickness ($\delta_{D}$) which employs both the streamwise turbulence intensity ($\sqrt{\overline{u^{2}}}$) and mean streamwise velocity (\ie akin to the diagnostic plot concept of \citealp{Alfredsson_Scaling_2011}) was introduced by \citet{Vinuesa_LengthScale_2016}: 
\begin{equation}
  \frac{\sqrt{\overline{u^{2}}}}{U}(z=\delta_{D})=0.02.
  \label{eq:dD}
\end{equation}
Here, $u$ represents instantaneous streamwise velocity fluctuations obtained through a conventional Reynolds decomposition (\ie $u=\tilde{U}-U$, where $\tilde{U}$ is the instantaneous streamwise velocity) and the subscript `$D$' denotes $\delta$ obtained through the diagnostic plot concept. 
The overline $(\overline{\cdot})$ indicates time averaging. 
The threshold prescribed in this method yields a TBL thickness, $\delta_{D}$, which was found to be equivalent to $\delta_{99}$ for the zero-pressure gradient (ZPG) TBL datasets tested \citep{Vinuesa_LengthScale_2016}. 
While this method has been successful in providing a more robust definition of mean TBL thickness for APG TBLs, the choice of threshold, which is tied to $\delta_{99}$ in this case, is still arbitrary (\ie it lacks physical interpretation). 
Additionally, because this definition relies on the decay of turbulence intensity in the far outer region towards the freestream turbulence level, the facility freestream turbulence level must be below the prescribed threshold (\ie $\scriptstyle \sqrt{\overline{u^{2}_{\infty}}} \textstyle /U_{\infty}<2\%$ from \citealp{Vinuesa_LengthScale_2016}) for this method to be applicable (or the threshold must be changed accordingly). 

Analogous to the decay of the streamwise turbulence intensity at the TBL edge, which was leveraged by \citet{Vinuesa_LengthScale_2016}, \citet{Wei_OuterScaling_2023} recently proposed using the asymptotic decay of the Reynolds shear stress ($\overline{uw}$) profile for defining the TBL thickness ($\delta_{uw}$), where $w$ is the wall-normal component of velocity fluctuations: 
\begin{equation}
  \overline{uw}(z=\delta_{uw})=0.01|\overline{uw}|_{\mathrm{max}}.
  \label{eq:duw}
\end{equation}
An advantage of this definition is that it is also translatable to other turbulent shear flows such as wakes or mixing layers to define their outer length scale \citep{Wei_OuterScaling_2023}. 
However, the threshold used in this method is again arbitrary (\ie not explained by any specific physical process within the TBL) and can be obscured by high levels of freestream turbulence. 
In addition, this method requires the simultaneous measurement of streamwise and wall-normal velocity fluctuations, which is more challenging to measure experimentally \citep{Lee_TurbulenceIntenisty_2016, Baidya_TurbulentStresses_2019}. 

\rev{While all of the above methods have been based on mean turbulence statistics, there are other methods, described below, which incorporate additional information about TBL physics by considering instantaneous variations in the local TBL thickness. 
The TNTI of the TBL, as described earlier, can be simulated or measured instantaneously using methods such as particle imaging velocimetry (PIV), to obtain two- or three-dimensional interfaces which represent the instantaneous TBL thickness (depending on the simulation or type of PIV used). 
The instantaneous coordinates of these measured interfaces (\egg $I=[x_{i},z_{i}]$ for a two-dimensional interface) can be ensemble averaged to generate a probability distribution, $P(x_{i},z_{i})$, which stochastically describes the TNTI location (\ie the local TBL thickness). 
A probability density function for the wall-normal height of the TNTI can then be found by summing this probability distribution across the streamwise direction, 
\begin{equation}
  \mathrm{p.d.f.}(z_{i})=\sum_{x}\frac{P(x_{i}, z_{i})}{dz}.
  \label{eq:pdf}
\end{equation}
\noindent This probability density function can further be approximated as a normal distribution with a measurable mean ($Z_{i}$) and standard deviation ($\sigma_{i}$). 
By definition, the boundary layer thickness should represent the outermost boundary of the turbulent flow (\ie the maximum height of the TNTI) beyond which only fully non-turbulent flow exists \citep{Chauhan_TNTI_2014}. 
Following \citet{Chauhan_TNTI_2014}, the properties of the probability density function ($Z_{i}$ and $\sigma_{i}$) can be used to estimate the highest wall-normal location the TNTI is expected to reach, on average, which can be considered as a surrogate of the TBL thickness, such that
\begin{equation}
  \delta_{\mathrm{TNTI}} = Z_{i} + 3\sigma_{i}.
  \label{eq:dtnti}
\end{equation}
While this method is related to important TBL physics, the choice of definition for the TBL thickness \citep[\egg three standard deviations above the mean TNTI height, following][]{Chauhan_TNTI_2014}, adds subjectivity to this method. 
However, this choice of definition, specifically the number of standard deviations, is not interpreted to be a free parameter like the thresholds used in previous methods.} 
Proper detection of the TNTI is also a highly active topic of research \citep{Reuther_TNTI_2018} and requires advanced/well-resolved measurement techniques/analysis \citep{Borrell_PropertiesTNTI_2016, Zecchetto_UniversalityTNTI_2021,lindic2025} in order to implement this method effectively (which is often not the case for large-scale experimental datasets). 
On the other hand, there are numerous methods to quantify intermittency (which is related statistically to properties of the TNTI) for conventional measurement techniques \citep[\egg hot-wire anemometry][]{Hedley_Intermittent_1974, De_Wavelet_2023} however, these methods also suffer from uncertainties and ambiguities emerging from the use of thresholds. 

Other methods for quantifying the TBL thickness have also been proposed, relying on quantities which are arguably even more complex and/or costly to obtain. 
Examples include the moment method \citep{Weyburne_Mathematical_2006}, methods based on mean vorticity \citep{Coleman_VorticityThickness_2018}, methods based on mean shear \citep{Vinuesa_LengthScale_2016}, or local reconstruction of the inviscid mean velocity profile \citep{Griffin_BLThickness_2021}. 
While these definitions can be physically insightful, the primary drawback is the requirement of significantly more advanced experimental techniques and/or simulations (with sufficient resolution in the outer region) for accurate application. 
This also means it would likely not be possible to retroactively apply these definitions to older, well-established datasets, where conventional techniques were used, for comparison. 

It should also be noted that many other characteristic `outer' length scales have been proposed in order to characterise the state of the TBL or to test the self-similarity of turbulence statistics (\egg the displacement and momentum thicknesses; \citealp{Schlichting_BLTheory_1955}, and various so called `mixing layer' scales; \citealp{Schatzman_ShearScaling_2017, Maciel_OuterScales_2018}). 
However, these length scales do not necessarily describe the outer edge of the boundary layer, which remains our primary focus. 
This study aims to propose a phenomenological definition for the mean TBL thickness that is independent of any thresholds, and can be applied retroactively to past single-point datasets irrespective of their canonical/non-canonical nature. 


\section{Experimental and numerical datasets}
\label{sec:data}

A set of experimental and numerical TBL datasets covering a broad range of Reynolds numbers, measurement techniques, and non-canonical conditions have been assembled and analysed here to compare the various definitions of the TBL thickness in the literature (summarised in table~\ref{tab:deltas}), as well as a new definition that will be formally proposed in $\S$~\ref{sec:definition}. 
The key details of these previously published and well-established datasets have been documented in $\S$~\ref{sec:exps_published}, with their parameters of interest also summarised in table~\ref{tab:exps_published} for reference. 
\rev{The ``+” superscript will be used throughout this paper to denote normalization by viscous velocity ($U_{\tau}$), length ($\nu/U_{\tau}$), and time ($\nu/U_{\tau}^2$) scales, where $\nu$ is the kinematic viscosity of air and $U_{\tau}$ is the mean friction velocity.} 
But first, we provide particular emphasis on a recent set of large-scale experiments \citep{Marusic_APGTNTI_2024,Lozier_APG_2024} conducted at the recently modified large Melbourne wind tunnel \citep{Deshpande_APG_2023}. 

\begin{table}
  \begin{center}
  \setlength{\tabcolsep}{10pt}
  \begin{tabular}{cccccc}
       Name & Type & Symbols & $\re$ & $\beta$ & $\ks$ \\ \hline \hline
       MELB1 & PIV & \ding{58} {\color{red}\ding{58}} & $7500$ & $0\ \&\ 1.4$ & \\
       MELB2 & HW & $\medblacksquare$ {\color{red}$\medblacksquare$} $\medblackcircle$ {\color{red}$\medblackcircle$} & $4500\rightarrow8000$ & $0\rightarrow1.5$ & \\
       MELB3 & PIV & \ding{58} {\color{mgreen}\ding{58}} & $6500\rightarrow12100$ & & $0\ \&\ 64$ \\
       MELB4 & HW & {\color{mgreen}$\medblacksquare$} & $3000\rightarrow29000$ & & $22\rightarrow155$ \\ \hline
       USNA1 & LDV & {\color{blue}$\medblacktriangleup$} $\medblacktriangleup$ {\color{red}$\medblacktriangleup$} & $300\rightarrow1900$ & -$1.0\rightarrow6.6$ & \\
       USNA2 & LDV & {\color{mgreen}$\medblacktriangleup$} & $600\rightarrow4700$ & & $32\rightarrow254$ \\
       USNA3 & LDV & {\color{mpurple}$\medblacktriangleup$} {\color{mgreen}$\medblacktriangleup$} {\color{myellow}$\medblacktriangleup$} & $600\rightarrow4700$ & -$0.7\rightarrow1.9$ & $30\rightarrow787$ \\ \hline
       UPM & DNS & & $1300\rightarrow2000$ &  & \\
       KTH & LES & & $2000$ & & \\ \hline
  \end{tabular}
  \caption{\rev{Details of datasets used in the current analysis. Blue, black and red symbols denote smooth-wall FPG, ZPG and APG cases, respectively. Magenta, green and yellow symbols denote rough-wall FPG, ZPG and APG cases, respectively. Arrows indicate many cases have been considered which cover the full parameter range listed.}}
  \label{tab:exps_published}
  \end{center}
\end{table}

\subsection{Recent large-scale experiments}
\label{sec:exps_melb}

\begin{figure}
\captionsetup{width=1.00\linewidth}
\begin{center}
\includegraphics[width=1.00\textwidth]{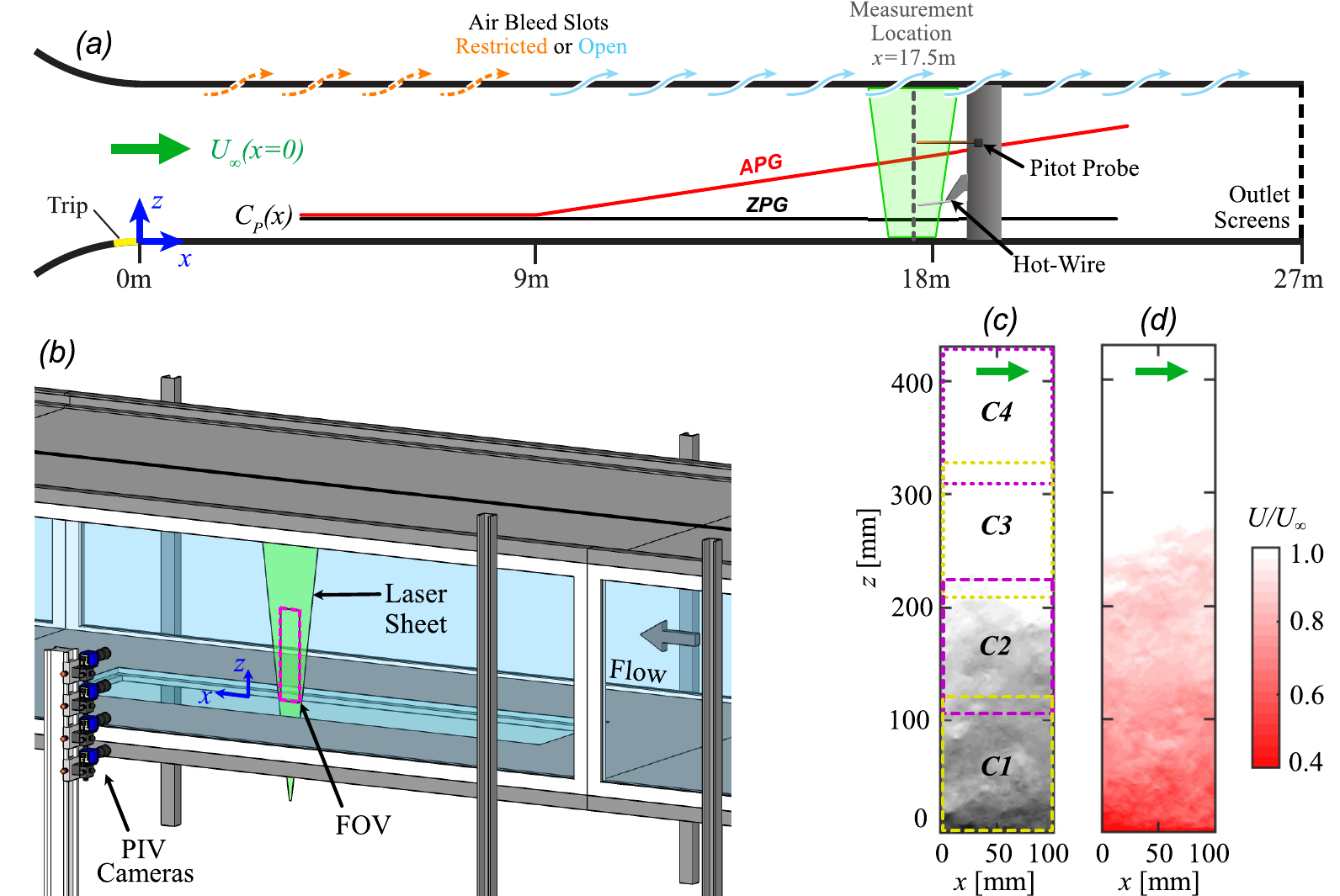}
\end{center}
\caption{(\textit{a}) Schematic of modified Melbourne large wind tunnel facility, adapted from \citet{Deshpande_APG_2023}. (\textit{b}) Schematic of PIV setup adapted from \citet{Marusic_APGTNTI_2024}. Snapshots of instantaneous streamwise velocity for (\textit{c}) ZPG and (\textit{d}) APG cases across the full TBL, made possible by stitching individual flow fields from the four PIV cameras (C1-C4).}
\label{fig1}
\end{figure}

For these recent datasets, we experimentally investigated moderately-strong APG TBLs at high Reynolds numbers using two measurement techniques, under matched conditions. 
This is made possible by recent modifications of the large Melbourne wind tunnel test section \citep{Deshpande_APG_2023}, shown schematically in figure~\ref{fig1}(\textit{a}). 
Low-porosity screens affixed to the outlet are used to raise the test section static pressure, while air bleed slots along the ceiling are opened (solid arrows in figure~\ref{fig1}\textit{a}) or restricted/closed (dashed arrows in figure~\ref{fig1}\textit{a}) to create a user controlled streamwise pressure gradient profile. 
For each case, the inlet unit Reynolds number ($U_{\infty}(x=0)/\nu=8.7E5$~m$^{-1}$) was held constant. 
In figure~\ref{fig1}(\textit{a}), the streamwise profiles of the pressure coefficient, $C_{P}(x) = 1-U^2_{\infty}(x)/U_{\infty}^{2}(x = 0)$, measured for two pressure gradient cases, one nominally ZPG and one with a mild APG, are overlaid for reference. 

A single hot-wire sensor was used in the first experiments (referred to as MELB2) to measure time-resolved velocity statistics across the TBL at the selected measurement location of $x=17.5$~m. 
The sensor was made in-house with a diameter of $d=2.5$~\textmu m and a nominal length of $l=0.5$~mm ($l^{+} \approx 11$). 
The hot-wire sampling frequency was $f_{s}=50$~kHz ($t^{+} \approx 0.3$) and the total sampling time ($T_{s}$) was set such that $T_{s}U_{\infty}(x)/\delta_{99}>20000$ for each case, to reach a reasonable level of statistical convergence. 
In each experiment, two independent profiles were acquired with unique wall-normal resolutions (\ie unique spacings between wall-normal measurement locations). 
The first profile followed a traditional logarithmic spacing, with $44$ total wall-normal measurement locations beginning near the wall and terminating in the freestream (square symbols in table~\ref{tab:exps_published}). 
In contrast, the second profile had $30$ linearly spaced wall-normal measurement locations restricted to the far outer region (${0.9\lesssim z/\delta_{99} \lesssim 1.4}$) leading to more data points near the TBL edge compared to the traditional profile (circle symbols in table~\ref{tab:exps_published}). 
The combination of these profiles then allows for the evaluation of conventional measurement practices, but also provides highly resolved measurements near the TBL edge for comparison of the various TBL thickness definitions. 
Calibration of the hot-wire probe was performed before and after each experiment to account for ambient drift over the long measurement duration, however hot-wire drift was confirmed to be negligible in these experiments. 
Due to the optical access at this streamwise measurement location, the friction velocity was obtained directly from oil-film interferometry for each case. 
Further details of the hot-wire measurements can be found in \cite{Marusic_APGTNTI_2024}. 

Complementing the hot-wire measurements, high-resolution PIV measurements (referred to as MELB1) were also conducted to capture detailed velocity fields across the entire boundary layer for both the ZPG and APG TBLs centred about a matched streamwise location of $x \approx 17.5$~m, as indicated in figure~\ref{fig1}(\textit{a}). 
The setup utilised four vertically staggered Imager CX-25 cameras (figure~\ref{fig1}\textit{b}) with $5312\times4608$ pixel complementary metal-oxide semiconductor (CMOS) sensors and Tamron SP AF 180~mm macro lenses set at $f/11$, achieving a digital resolution of 22~\textmu m/pixel. 
The positioning of the cameras ensured complete TBL coverage, with $\delta_{99}$ for both the ZPG and APG TBLs lying within the field of view of the middle cameras (figure~\ref{fig1}\textit{c}). 
\rev{A dual-pulse Nd:YAG laser (InnoLas SpitLight Compact PIV 400) with a 2~mm thick laser sheet (corresponding to a viscous-scaled thickness of $\sim 40$-$50$) illuminated the flow which was seeded with 1–2~\textmu m particles, while synchronisation was handled by a programmable timing unit (PTU X, LaVision GmbH) via DaVis 10.1 software.} 
The final stitched field-of-view (FOV; figure~\ref{fig1}\textit{b}) measured $104 \times 441$~mm\textsuperscript{2} in the streamwise and wall-normal directions ($x \times z$).

A 2-D dot target was used for camera calibration, and a minimum intensity subtraction technique enhanced image quality. 
\rev{Multi-pass cross-correlation was applied with a final interrogation window size of $24 \times 24$~pixels ($0.53 \times 0.53$~mm\textsuperscript{2}) and 50\% overlap, yielding viscous-scaled spatial resolutions of $18 \times 18$ for the ZPG and $11 \times 11$ for the APG case ($x^{+} \times z^{+}$).} 
A sample of the instantaneous streamwise velocity field for both the ZPG and APG case are given in figures~\ref{fig1}(\textit{c,d}) respectively. 
Further details of the PIV setup can be found in \citet{Marusic_APGTNTI_2024} and \citet{lindic2025}, which also give details of the specific TNTI detection methodology adopted for the present study, and the challenges associated with adopting other methodologies for experimental datasets. 

\subsection{Published datasets}
\label{sec:exps_published}

A set of published datasets were also considered, supplementing the current analysis with different experimental/numerical techniques and unique combinations of non-canonical effects. 
These datasets are described briefly below, with the relevant citations provided for further details. 

The datasets referred to as MELB3 and MELB4 in table~\ref{tab:exps_published} are from a series of previously published experimental studies documenting the effects of surface roughness on high friction Reynolds number (${Re_{\tau}\equiv U_\tau\delta_{99}/\nu}$) TBLs, which were conducted in the same facility as MELB1 and MELB2. 
In both of these studies, surface roughness was introduced by covering the entire bottom wall of the test section with a single sheet of sandpaper. 
The surface roughness is quantified by an equivalent sand grain roughness Reynolds number, $\ks=k_{S}U_{\tau}/\nu$. 
Specifically, the dataset MELB3 is a set of PIV measurements of ZPG smooth- and rough-wall TBLs documented in \citet{Squire_SmoothRoughPIV_2016}. 
While, the dataset MELB4 is associated with hot-wire measurements of ZPG rough-wall TBLs documented in \cite{Squire_SmoothRoughHW_2016}, both of which can be directly consulted for further details on the experimental setup. 

Datasets with the prefix USNA in table \ref{tab:exps_published} correspond to laser Doppler velocimetry (LDV) measurements conducted in the US Naval Academy water channel. 
Here, a range of pressure gradient conditions, starting from a favourable-pressure gradient (FPG) and ending with an APG, were introduced by adjusting four flat plates along the upper wall of the channel. 
Additionally, varying levels of surface roughness were introduced through interchangeable plates, which made up the bottom wall of the channel. 
Datasets USNA1, USNA2 and USNA3 respectively correspond to LDV measurements of smooth-wall pressure gradient TBLs \citep{Volino_SmoothPG_2020}, ZPG rough-wall TBLs \citep{Volino_RoughZPG_2022} and rough-wall pressure gradient TBLs \citep{Volino_SmoothRoughPG_2023}. 

Additionally, two numerical simulations of canonical TBLs were also considered, supplementing the current analysis with a wider range of measurement resolutions and flow conditions. 
Datasets referred to as UPM and KTH in table~\ref{tab:exps_published} correspond to a direct numerical simulation (DNS) and a well-resolved large-eddy simulation (LES) of a smooth-wall ZPG TBL, respectively. 
Details of the DNS and LES are documented in \citet{Sillero_ZPGDNS_2013} and \cite{EitelAmor_ZPGLES_2014}, respectively. 


\section{Definition of mean TBL thickness based on streamwise velocity skewness}
\label{sec:definition}

While developing this new definition for the TBL thickness, we considered various criteria in an effort to ensure that the new definition is practical, and broadly applicable. 
First, it is ideal for the new definition to be implementable when using conventional experimental measurement techniques (\ie single velocity component, single-point measurements), in addition to more advanced experimental measurement techniques and simulations. 
This would also ensure that the new method can be applied retroactively, on other well-established/published datasets. 
Second, the use of thresholds should be avoided, if possible, to reduce ambiguity and/or bias. 
And third, the new definition should be relatable back to meaningful TBL physics. 
In meeting these criteria, we aim to overcome the shortcomings of other methods established in the literature, as summarised in table~\ref{tab:deltas}. 
To that end, we propose the following definition: 
\begin{equation}
  \overline{\overline{u^{3}}}(z=\ds)=0,
  \label{eq:dS}
\end{equation}
\noindent where the local mean turbulent boundary layer thickness is defined as the wall-normal location where the skewness of streamwise velocity fluctuations, in the outermost region, changes sign from negative to positive. 
Here, the double overline denotes the appropriate normalisation by the variance of streamwise velocity fluctuations at the corresponding $z$-location (\ie~$\scriptstyle {\overline{\overline{u^{3}}} = \overline{u^{3}}/\sqrt{\overline{u^{2}}}^{3}}$). 
This is a conventional definition of skewness, and will be applied consistently throughout the current analysis. 
The skewness of streamwise velocity can be easily measured in conventional experiments, and the change of sign in the skewness profile in the outer region means no thresholds are imposed. 
It is noted that past studies have also analysed ${\overline{\overline{u^{3}}}}$ \egg to interpret the scaling of probability distribution functions or quantify non-linear triadic interactions \citep{duvvuri2015,lozier2024}. 

\subsection{Physical insights}
\label{sec:insights}

\begin{figure}
\captionsetup{width=1.00\linewidth}
\begin{center}
\includegraphics[width=0.90\textwidth]{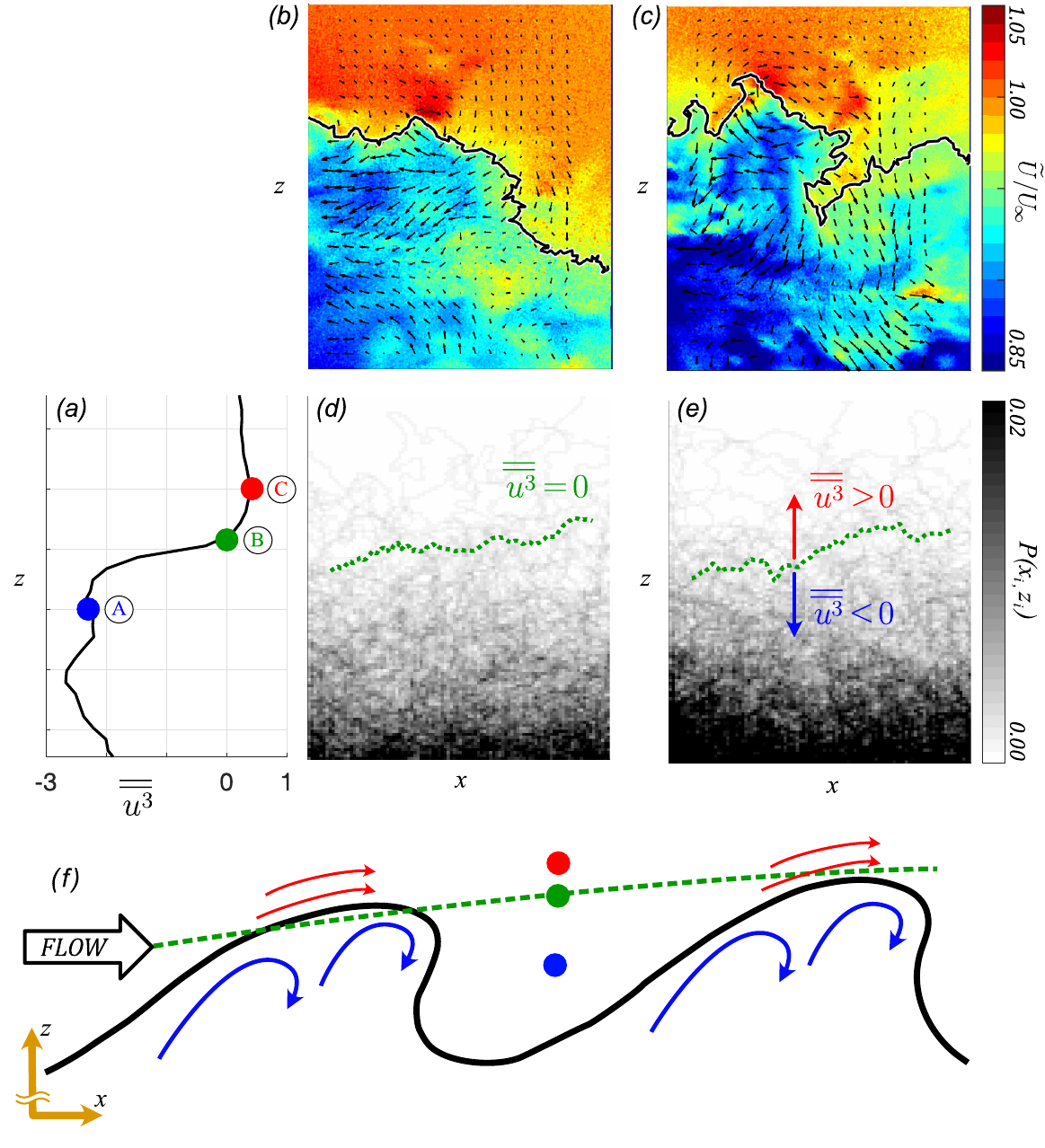}
\end{center}
\caption{\rev{(\textit{a}) Profile of streamwise velocity skewness in the outer region of a TBL. (\textit{b,c}) Instantaneous LKE interface (black lines) imposed on the instantaneous streamwise velocity field (colours) with instantaneous fluctuation vectors (arrows) overlaid. (\textit{d,e}) Probability distribution of the interface location in the outer region of a TBL with contours of zero skewness overlaid. Data are from the MELB1 (\textit{a,b,d}) ZPG and (\textit{c,e}) APG cases. (\textit{f}) Schematic of instantaneous flow phenomenology associated with the characteristic wall-normal variation of streamwise velocity as shown in (\textit{a,b,c})}.}
\label{fig2}
\end{figure}

An example of a representative skewness profile for a canonical TBL, demonstrating the significance of this sign change, can be seen in figure~\ref{fig2}(\textit{a}). 
\rev{The green symbol (and green dotted lines in figures~\ref{fig2}\textit{d-f}) indicates the wall-normal location where the skewness changes from negative to positive, labelled as `B'. 
The red and blue symbols (labelled `C' and `A') indicate the wall-normal locations of the outer positive and negative peaks in the skewness profile, respectively, which bound the zero-crossing. 
Figures~\ref{fig2}(\textit{b,c}) demonstrate the flow phenomenology associated with the skewness of streamwise velocity fluctuations. 
Here the background colour represents the instantaneous streamwise velocity magnitude and the arrows represent instantaneous fluctuations in the streamwise and wall-normal velocity. 
Following \citet{lindic2025}, the local turbulent kinetic energy (LKE, $\tilde{k}$) in a frame of reference moving with the freestream velocity, over a $3 \times 3$ grid is defined as, 
\begin{equation}
      \tilde{k}=100 \times \frac{1}{9 U_{\infty}^2} \sum_{m,n=-1}^1\left[\left(\tilde{U}_{m, n}-U_{\infty}\right)^2+\left(\tilde{W}_{m,n}-W_{\infty}\right)^2\right],
  \label{eq:LKE}
\end{equation}
\noindent and the instantaneous interface coordinates $[x_{i},z_{i}]$ can be defined by a contour of the LKE where $\tilde{k}(x_{i},z_{i})=0.2$, represented by solid black lines in figures~\ref{fig2}(\textit{b,c}).
Fluctuations above this interface are typically weak, and the instantaneous velocity is typically either equal to, or slightly greater than, the freestream velocity (${\tilde{U}\gtrsim U_\infty}$). 
The latter is attributed to a local acceleration of the streamwise velocity above the evolving interface/bulges. 
This acceleration was observed experimentally, over a small wall-normal extent above the LKE-defined interface, in approximately 40\% of snapshots for both ZPG and APG TBLs, like those shown in figures~\ref{fig2}(\textit{b,c}). 
In the other snapshots, the instantaneous velocity in the region above the interface was indistinguishable from the freestream velocity. 
Below the interface, there are strong turbulent fluctuations, and the instantaneous velocity is consistently lower than the freestream velocity. 
While the relative position of these velocity fluctuations/features, with respect to the interface, may vary with interface detection method (\egg LKE in this case; \citealp{lindic2025}) they are still expected to be highly correlated with the interface topology. 
Additionally, the commonality of these flow features between the ZPG and APG TBLs is not surprising given the qualitatively similar coherent flow structures/energy dynamics in their respective far outer regions \citep{lee2017,deshpande2024}. 
These flow features were also observed in a ZPG rough-wall TBL using the PIV dataset MELB3 from \citet{Squire_SmoothRoughPIV_2016}, though not shown here for conciseness.}

Figure~\ref{fig2}(\textit{f}) schematically relates the wall-normal variation in TBL flow features with the wall-normal profile of the skewness of streamwise velocity fluctuations. 
For instance, positive skewness (red) arises in a region where the flow is primarily freestream, with intermittent accelerations of the flow as the turbulent bulges pass below, through the far outer region. 
Alternatively, negative skewness (blue) appears in a region where the flow experiences freestream flow with intermittent, turbulent, low instantaneous velocity events, \ie events associated with the turbulent eddies within the TBL. 
Finally, there is a region in which the flow sharply transitions from the negative to positive skewness state with increasing wall-normal distance, creating a zero-crossing which is identified and used in \eqref{eq:dS} as the metric by which to estimate the boundary layer thickness, $\ds$. 
\rev{In figures~\ref{fig2}(\textit{d,e}) 2000 instantaneous LKE interfaces are ensembled averaged to create a two-dimensional probability distribution for the interface, $P(x_{i},z_{i})$, with an overlaid contour of locally zero-skewness (green dotted line). 
While interfaces can (infrequently) exceed the contour of zero-skewness, the zero-skewness contour acts as a nominal indicator of the expected uppermost extent of the interface, on average}. 
In this way the definition of skewness proposed here is phenomenologically similar to \eqref{eq:dtnti}, but does not rely on thresholds and can be readily applied to single-point measurements of the streamwise component of velocity (even retroactively). 
\rex{Additionally, following the theoretical framework of \citet{Phillips_Irrotational_1955}, the skewness of velocity fluctuations parallel to the mean TNTI (\ie in a rotated frame of reference accounting for TBL growth) were also found to be positive for $z \geq \ds$, akin to the skewness of streamwise velocity fluctuations considered here. 
This observation reaffirms that the characteristic velocity fluctuations in the far outer region of the TBL (described above) have a strong relationship with the local TNTI dynamics (rather than mean TNTI location alone). 
These results also suggest Phillips' model for velocity fluctuations outside the turbulent boundary should perhaps be tested relative to the \emph{instantaneous} TNTI, which is however beyond the scope of the present work.} 

\subsection{Attached eddy modelling}
\label{sec:AEM}

\begin{figure}
\captionsetup{width=1.00\linewidth}
\begin{center}
\includegraphics[width=1.00\textwidth]{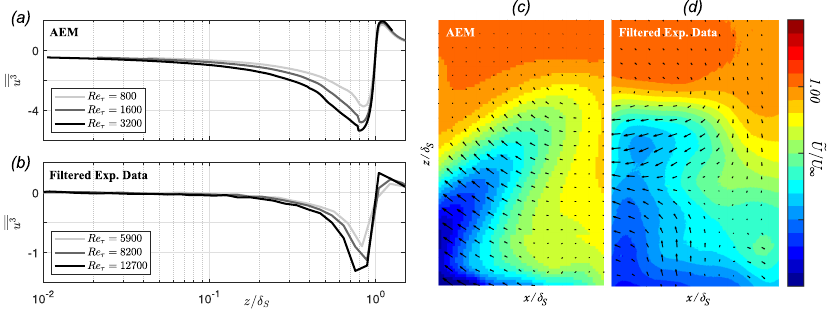}
\end{center}
\caption{\rex{(\textit{a}) Skewness profiles and (\textit{c}) a snapshot of the instantaneous velocity field from \citet{Deshpande_AEM_2021}. (\textit{b}) Filtered  skewness profiles from \citet{lozier2024}. (\textit{d}) Filtered PIV snapshot of the instantaneous velocity field from figure~\ref{fig2}(\textit{b}).}}
\label{figA1}
\end{figure}

\rex{The generic nature of these flow features can further be demonstrated by considering the simulations of \citet{Deshpande_AEM_2021} which incorporated empirically obtained geometric scaling laws into the classical attached eddy model (AEM) \citep{Perry_WallTurbulence_1982, MarusicMonty2019} to simulate the coherent large-scale flow structures typical of ZPG TBLs. 
Specifically, this model consisted of a data-driven distribution of three-dimensional `packets' of $\Lambda$-eddies, representing geometrically self-similar eddies and $\delta$-scaled large-scale motions (LSMs), but did not include information related to scale-specific non-linear interactions \citep{duvvuri2015, lozier2024}. 
A snapshot of the instantaneous velocity fields from one such simulation can be seen in figure~\ref{figA1}(\textit{c}) along with the skewness profiles from a set of simulations with varying $\re$ in figure~\ref{figA1}(\textit{a}). 
Key characteristics of both the instantaneous velocity field from figure~\ref{fig2}(\textit{b}) (\egg a slight acceleration above the largest eddies with lower-velocity motions below) and the skewness profile from figure~\ref{fig2}(\textit{a}) (\egg the distinct positive and negative peaks) can also be seen in these AEM results, reaffirming the phenomenological description behind the $\ds$ definition, and its broad applicability. 
For comparison, scale-dependent filtering \citep[$u_{\mathrm{filtered}} = u \lbrack T^{+} > 350 \rbrack $;][]{lozier2024} was also applied to relevant experimental data in order to isolate the contributions of LSMs to the measured instantaneous velocity fields and skewness profiles shown in figure~\ref{fig2}. 
Filtered skewness profiles from \citet{lozier2024} and a filtered PIV snapshot reproduced from figure~\ref{fig2}(\textit{b}) are shown in figures~\ref{figA1}(\textit{b}) and (\textit{d}), respectively. 
Critically, the key characteristics of the instantaneous velocity field and skewness profiles are retained after filtering, and agree well with the AEM results. 
These results collectively reinforce the dependence of the skewness profile on the influence of LSMs (which grows with $\re$), and support the physical insights which led to the establishment of this new $\ds$ definition. 
However, highly-resolved DNS are also recommended to investigate this relationship between TNTI dynamics and skewness in greater detail \citep{Borrell_PropertiesTNTI_2016, Zecchetto_UniversalityTNTI_2021}.} 

\subsection{Comparison with other definitions}
\label{sec:compare}

\begin{figure}
\captionsetup{width=1.00\linewidth}
\begin{center}
\includegraphics[width=1.00\textwidth]{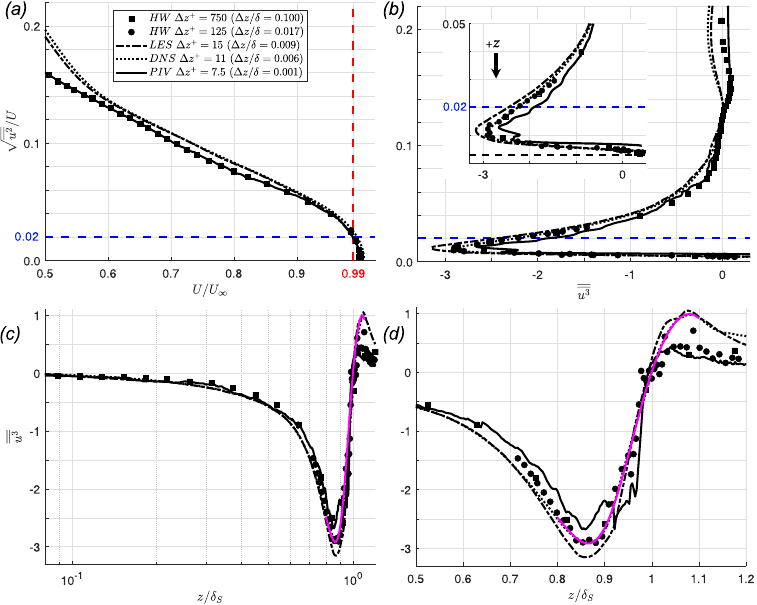}
\end{center}
\caption{\rev{Comparison of experimental and numerical ZPG TBL statistics with varying wall-normal resolutions. (\textit{a}) Diagnostic style plot used to find $\delta_{D}$ following the methodology of \citet{Vinuesa_LengthScale_2016} (analogous to $\delta_{99}$). (\textit{b}) Relationship between turbulence intensity and skewness of streamwise velocity fluctuations in the outer region of the TBL. Wall-normal profiles of skewness normalised by $\ds$ in (\textit{c}) logarithmic scaling and (\textit{d}) linear scaling. The magenta curve represents a generalised form of the normalised skewness profile \eqref{eq:FM} fit to DNS data from \cite{Sillero_ZPGDNS_2013}.}}
\label{fig3}
\end{figure}

To compare past $\delta$ definitions with this new $\ds$ definition, figure~\ref{fig3}(\textit{a}) shows a diagnostic style plot \citep{Alfredsson_Scaling_2011, Vinuesa_LengthScale_2016} with an ensemble of published smooth-wall ZPG (\ie canonical) TBL datasets, from both numerical and experimental studies, encompassing a broad range of wall-normal resolutions. 
The vertical dashed red line shows the threshold where the mean velocity is equal to 99\% of the freestream velocity (\ie $\delta_{99}$), while the horizontal blue dashed line shows the turbulence intensity threshold used to find the boundary layer thickness $\delta_{D}$ \citep{Vinuesa_LengthScale_2016}. 
All the ZPG datasets appear to pass through the intersection of these two thresholds, confirming that, $\delta_{99}=\delta_{D}$ for ZPG TBLs, consistent with \cite{Vinuesa_LengthScale_2016}. 
In fact, this relationship was also found to hold for the non-canonical datasets considered in this study (see table~\ref{tab:exps_published}) and as such, we will use \eqref{eq:dD} to find $\delta_{99}$, for consistency, dropping the $\delta_{D}$ terminology from here on out. 
Figure~\ref{fig3}(\textit{a}) also demonstrates some key limitations of these methods, associated in particular with the asymptotic nature of first- and second-order statistics in ZPG TBLs. 
For instance, as described in $\S$~\ref{sec:intro}, the $\delta_{99}$ definition is predicated on the assumption that the mean velocity profile monotonically approaches the freestream velocity with increasing wall-normal distance ($z$) through the outer region. 
While this is true for canonical TBLs (see figure~\ref{fig3}(\textit{a}), this behaviour has been shown to differ in the case of non-canonical TBLs, such as APG TBLs \citep{Vinuesa_LengthScale_2016, Griffin_BLThickness_2021}. 
Similarly, the diagnostic plot method also relies on the streamwise turbulence intensity decaying towards a relatively low freestream turbulence intensity level as the wall-normal distance increases through the outer region. 
If the freestream turbulence intensity is near or above the originally prescribed threshold \citep[2\% from][]{Vinuesa_LengthScale_2016}, determining the boundary layer thickness may require modification of the threshold accordingly. 
Additionally, the first- and second-order statistics do not capture all the important physics happening in the outer region, which is evident on comparing figures~\ref{fig2} and \ref{fig3}(\textit{a}). 

\rev{To further demonstrate this, the vertical axis in figure~\ref{fig3}(\textit{b}) shows the streamwise turbulence intensity, like figure~\ref{fig3}(\textit{a}), however the normalised mean streamwise velocity on the horizontal axis has been replaced with the streamwise velocity skewness. 
As the streamwise turbulence intensity decreases (corresponding to increasing wall-normal distance indicated by the vertical arrow), significant and non-monotonic variations in skewness are observed. 
This remains true even as the streamwise turbulence intensity approaches the freestream turbulence intensity level, $\scriptstyle \sqrt{\overline{u^{2}_{\infty}}} \textstyle /U_{\infty}=0.003$ for the experimental datasets shown here (marked by horizontal black dashed line in figure~\ref{fig3}(\textit{b}) inset). 
This indicates there are still significant skewness contributing events occurring near the TBL edge, as described in figure~\ref{fig2}, which give rise to the unique profile of skewness seen in the far outer region of the TBL, and which can not be captured by first- or second-order statistics alone (figure~\ref{fig3}\textit{a}).} 
To that end, the wall-normal profiles of skewness for each case are shown in figures~\ref{fig3}(\textit{c,d}). 
Here, the wall-normal distances have been normalised using $\ds$ following \eqref{eq:dS}. 
From these figures we can see that, for smooth-wall ZPG TBLs, the skewness profiles all agree reasonably well both above and below the point of $z=\ds$. 
Visually identifying the zero-crossing in the skewness profiles here is intuitive. 
However, for the experimental datasets, noise and/or large wall-normal spacing between points can make determination of the outer zero-crossing of the skewness profile relatively more challenging as compared to the numerical datasets. 
To assist with this effort, we propose fitting a model equation to the outer region of the skewness profile, to extract the location of the zero-crossing more consistently. 
The magenta lines in figures~\ref{fig3}(\textit{c,d}) represent a Fourier model of the skewness profile in the far outer region of the TBL, given by 
\begin{equation}
  \overline{\overline{u^{3}}} = a_{0} + \sum_{n=1}^{3} a_{n}\cos(n\omega\frac{z}{\ds}) + b_{n}\sin(n\omega\frac{z}{\ds}),
  \label{eq:FM}
\end{equation}
which has been fitted to the DNS skewness profile within our region of interest, $0.8\leq z/\ds \leq 1.1$. 
The resulting coefficients were: $a_{0}=-1.06, ~a_{1}=0.67, ~a_{2}=-0.01, ~a_{3}=0.08, ~b_{1}=1.82, ~b_{2}=0.23, ~b_{3}=-0.05$ and $\omega =12.73$. 
The ZPG TBLs considered here all agree well with this Fourier model, and going forward we will consider this Fourier model as a generalised representation of the skewness profile in the far outer region of a smooth-wall ZPG TBL. 
In this way we can fit all datasets considered here to this model in order to extract $\ds$, as done in figures~\ref{fig3}(\textit{c,d}), and all subsequent analysis. 
Additionally, this method involving the Fourier model will be directly compared with simple linear interpolation in $\S$~\ref{sec:applicability}. 

Interestingly, it can be noted in figure~\ref{fig3}(\textit{d}) that the positive peak in the skewness profile, beyond the boundary layer thickness ($z/\ds>1$), is lower in magnitude for the experimental datasets as compared to the numerical datasets (and subsequently the Fourier model). 
This difference is likely a result of Gaussian freestream turbulence which, in a region with some skewness contributing events (positive or negative), will tend to bring the measured skewness towards zero. 
However, even with differences in the peak amplitudes, the experimental data still shows a shape which is consistent with the numerical data and the Fourier model. 
This also confirms that Gaussian freestream turbulence and/or measurement noise should not change the location of the zero-crossing (and consequently $\ds$) since it does not contribute strong positive or negative skewness to the velocity signal. 
\rev{Along similar lines, the agreement of the various experimental profiles also indicates that they have reasonable statistical convergence, which is important when considering higher-order statistics such as skewness. 
However, it is noteworthy that while the hot-wire measurements achieve convergence temporally (\ie through long sampling times), the PIV measurements achieve a similar level of convergence through spatial and ensemble averaging.} 
Additionally, figure~\ref{fig3}(\textit{d}) demonstrates the effect of measurement wall-normal resolution ($\Delta z$) on the accuracy of resolving the skewness profile and its zero-crossing. 
The experimental and numerical datasets with small wall-normal resolutions (\ie $\Delta z^{+}\leq 125$ and $\Delta z/\ds \leq 0.017$) appear to follow the Fourier model well. 
Note that the hot-wire dataset with the poorest spatial resolution (square symbols) is typical of experiments with logarithmically spaced measurement points, where the distance between measurement points is large in the far outer region. 
However, there are still multiple points within the region of interest (\ie $0.8 \leq z/\ds \leq 1.1$) and fitting these points to the model results in a good estimate of $\ds$, consistent with that obtained by fitting the model to better resolved hot-wire statistics. 
Next, we apply our new definition of the boundary layer thickness \eqref{eq:dS} and compare it with past definitions used in the literature. 
Later, in $\S$~\ref{sec:applicability} we also demonstrate the robustness of our $\ds$ determination method (as described above) on a range of previously-published, single-point experimental datasets in both canonical and non-canonical TBLs. 

\begin{figure}
\captionsetup{width=1.00\linewidth}
\begin{center}
\includegraphics[width=1.00\textwidth]{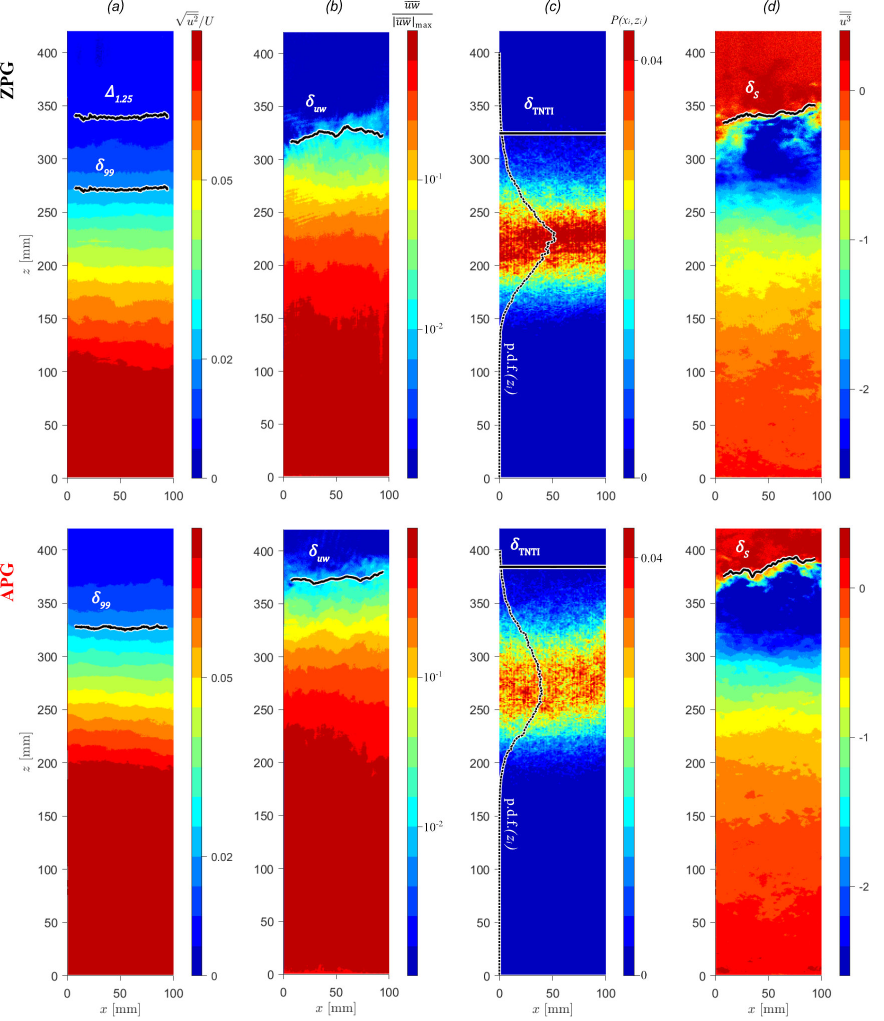}
\end{center}
\caption{\rev{Two-dimensional fields of relevant statistics from PIV measurements of (\textit{top}) ZPG and (\textit{bottom}) APG TBLs with solid contours of the TBL thickness overlaid based on (\textit{a}) $\delta_{99}$, $\Delta_{1.25}$, (\textit{b}) $\delta_{uw}$, (\textit{c}) $\delta_{\mathrm{TNTI}}$, and (\textit{d}) $\ds$ definitions. The black dotted lines overlaid in (\textit{c}) represent p.d.f.s of the TNTI height \eqref{eq:pdf}.}}
\label{fig4}
\end{figure}

Plots in figure~\ref{fig4} compare our new definition of the TBL thickness with other commonly used definitions of the boundary layer thickness (as summarised in table~\ref{tab:deltas}) for both ZPG and APG TBLs. 
It is clear that $\ds$ (shown in figure~\ref{fig4}\textit{d}) is larger (farther from the wall) than $\delta_{99}$ (shown in figure~\ref{fig4}\textit{a}) for both the ZPG and APG cases. 
However, for the ZPG case, $\ds$ agrees reasonably well with $\Delta_{1.25}$ (\ie $1.25\delta_{99}$, figure~\ref{fig4}\textit{a}), consistent with the relationship established in \citet{Baxerres_PGBoundaryLayers_2024} using a composite profile of the mean velocity. 
In the case of APG TBLs, however, there is currently a lack of universal composite profile for the outer region of the mean velocity profile which could be used to find an equivalent $\Delta$ parameter. 
In both the ZPG and APG cases, $\ds$ is also larger than $\delta_{uw}$ (shown in figure~\ref{fig4}\textit{b}) by $\approx 11\%$, but it is comparable, suggesting that a slightly different threshold of $\overline{uw}$ may result in the same boundary layer thickness. 
Similarly, the boundary layer thickness found using the p.d.f. of the TNTI height ($\delta_{\mathrm{TNTI}}$) agrees well with $\ds$ ($\approx 4\%$ difference), for both the ZPG and APG TBLs shown in figure~\ref{fig4}(\textit{c}). 
These results show that this new definition of boundary layer thickness, which is motivated by characteristic TBL physics (depicted in figure~\ref{fig2}\textit{f}), yields results comparable to previously used definitions, but without the use of thresholds. 

\begin{figure}
\captionsetup{width=1.00\linewidth}
\begin{center}
\includegraphics[width=1.00\textwidth]{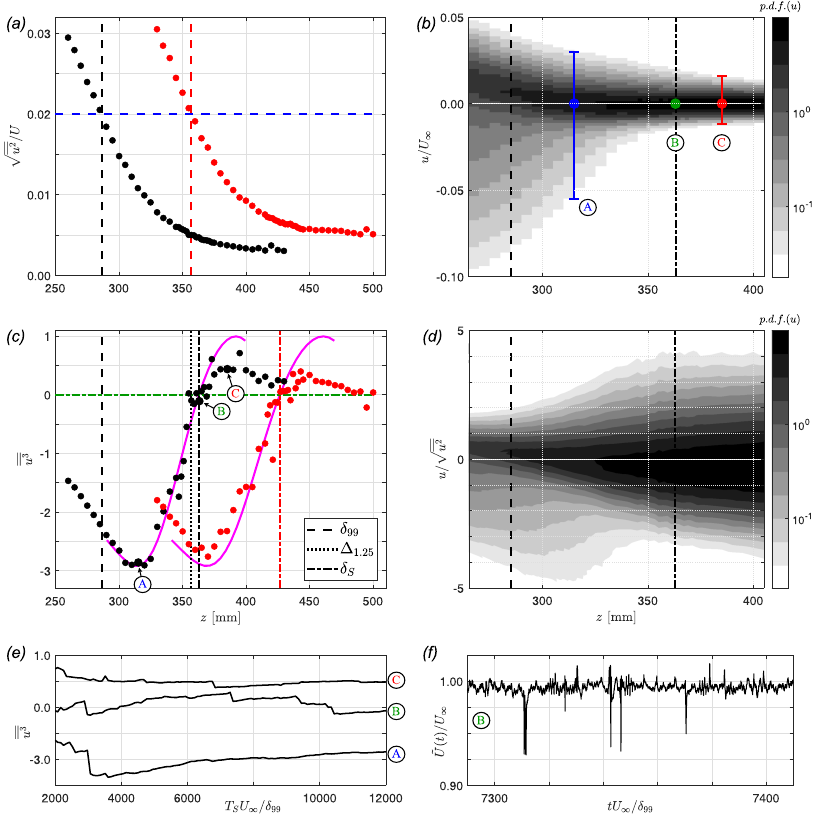}
\end{center}
\caption{\rev{Wall-normal profiles of (\textit{a}) turbulence intensity and (\textit{c}) skewness of streamwise velocity from high-resolution hot-wire measurements of ZPG (in black) and APG (in red) TBLs. The p.d.f. of streamwise velocity fluctuations, as a function of wall-normal distance, in the outer region of the ZPG TBL normalised by (\textit{b}) the freestream velocity and (\textit{d}) local turbulence intensity. (\textit{e}) Effect of sampling time on the magnitude of skewness measured at wall-normal locations corresponding to the negative peak, zero-crossing, and positive peak in the ZPG TBL skewness profile (labelled `A', `B' and `C' respectively). \rex{(\textit{f}) Time series excerpt of the instantaneous velocity measured near the zero-crossing.}}}
\label{fig5}
\end{figure}

\rev{To complement the results presented in figure~\ref{fig4}, hot-wire measurements of the outer region, with a dense linear wall-normal spacing, were conducted for both the ZPG (black) and APG TBLs (red), at conditions matched to the PIV experiments, and the results are shown in figure~\ref{fig5}. 
These experiments are used to further examine the skewness profile in the outer region and demonstrate how the new $\ds$ definition can be applied to conventional single-point measurements (and retroactively to previously acquired datasets). 
Figure~\ref{fig5}(\textit{a}) shows the normalised turbulence intensity for both cases as a function of wall-normal distance, with $\delta_{99}$ being estimated using \eqref{eq:dD} (conventional threshold shown as a dashed blue line). 
Similarly, figure~\ref{fig5}(\textit{c}) shows the wall-normal profile of skewness for both cases, with $\ds$ estimated by fitting the skewness profiles with the Fourier model (equation~\ref{eq:FM}, zero skewness is highlighted with a dash dotted green line). 
For the ZPG case, $\ds$ and $\Delta_{1.25}$ are compared directly, demonstrating good agreement once again. 
Additionally, the relative error in the skewness measurement at critical wall-normal locations in the profile can be estimated by examining how the magnitude of skewness changes with increasing sampling time, as shown in figure~\ref{fig5}(\textit{e}). 
Select measurement points corresponding to the negative peak, zero-crossing, and positive peak in the skewness profile have been labelled as `A', `B' and `C' respectively following figure~\ref{fig2}. 
For the point aligned with the negative peak (`A'), the skewness magnitude appears to be monotonically converging for sampling times greater than $\approx 8000$ large-eddy turnover times ($\delta_{99}/U_{\infty}$). 
However, for the points near the zero-crossing and aligned with the positive peak, the skewness magnitude is highly sensitive to infrequent but significant skewness-contributing events resulting in a relatively high uncertainty even after a cumulative sampling time of $12000$ large-eddy turnovers. 
\rex{To highlight this, an excerpt of the instantaneous velocity time series, which aligns with a sharp jump in skewness magnitude in figure~\ref{fig5}(\textit{e}), is shown in figure~\ref{fig5}(\textit{f}). 
Even with an accumulated sampling time of nearly $7300$ large-eddy turnovers, adding this additional sample of around $100$ turnover times, which contains several large fluctuations (\ie skewness-contributing events), results in the significant jump in skewness magnitude observed in figure~\ref{fig5}(\textit{e}). 
This time series excerpt also reaffirms that, near the zero-crossing, the instantaneous velocity can intermittently exceed the freestream velocity magnitude and/or drop significantly below the local mean consistent with the observations/model presented in \S~\ref{sec:insights} and figure~\ref{fig2}.} 
Despite a persistent uncertainty in skewness around the zero-crossing, the estimation of $\ds$ is accurate when fitting these profiles with the Fourier model, as seen in figure~\ref{fig5}(\textit{b}), as the fitting method considers \emph{many} points between the negative and positive peaks, which mitigates this uncertainty. 
It should also be reiterated that while the hot-wire measurements shown here achieve reasonable statistical convergence temporally (\ie through sufficiently long sampling times), other measurements can alternatively achieve statistical convergence through spatial and/or ensemble averaging. 
The p.d.f.s of streamwise velocity fluctuations across the outer region are also shown in figures~\ref{fig5}(\textit{b,d}). 
In figure~\ref{fig5}(\textit{b}), the velocity fluctuations are normalised by the freestream velocity, with the extent of distribution at the positive and negative peaks (`A' and `C') highlighted to bring out the `tailedness' responsible for the skewness magnitude. 
In figure~\ref{fig5}(\textit{d}), the fluctuations have been normalised by the turbulence intensity to explicitly bring out the negative and positive peaks, and demonstrate the recovery of $u$-fluctuations to the expected Gaussian distribution in the freestream. 
It should also be highlighted that while the distribution clearly transitions from negative to positive, the distribution is \emph{not} Gaussian at locations near the zero-crossing.
The same has also been confirmed based on the locally large kurtosis magnitudes measured at and near the zero-crossing, which are not been shown here for brevity.} 


\section{Applicability of new definition to previously published datasets}
\label{sec:applicability}

\begin{figure}
\captionsetup{width=1.00\linewidth}
\begin{center}
\includegraphics[width=0.90\textwidth]{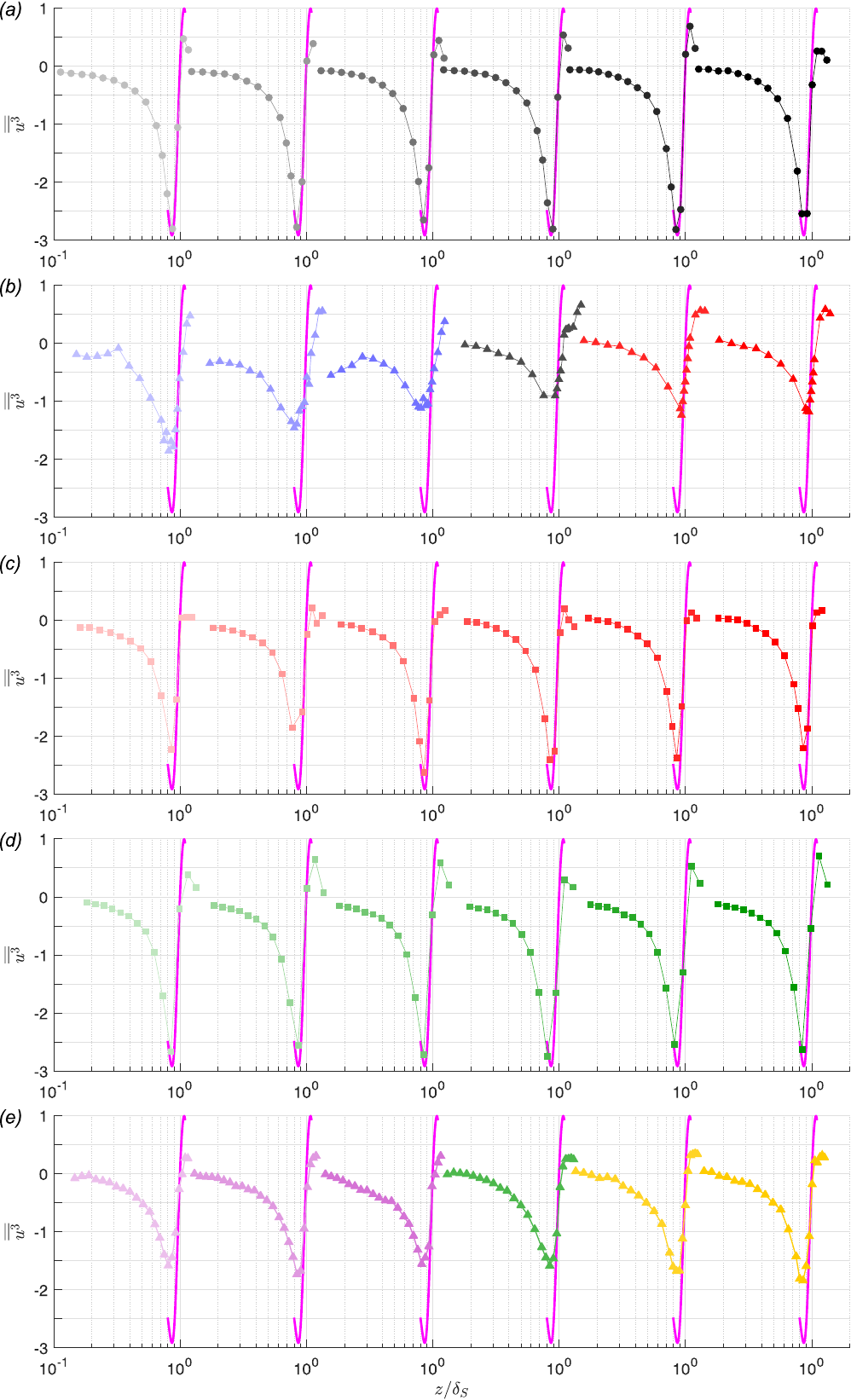}
\end{center}
\caption{Select normalised wall-normal profiles of streamwise velocity skewness from (\textit{a}) MELB2, (\textit{b}) USNA1, (\textit{c}) MELB2, (\textit{d}) MELB4 and (\textit{e}) USNA3.}
\label{fig6}
\end{figure}

\begin{figure}
\captionsetup{width=1.00\linewidth}
\begin{center}
\includegraphics[width=1.00\textwidth]{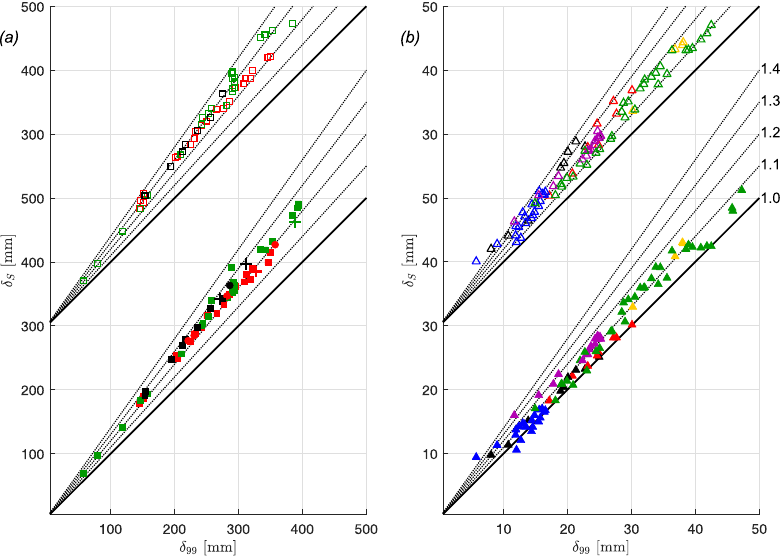}
\end{center}
\caption{Comparison of $\ds$ and $\delta_{99}$ for select (\textit{a}) Melbourne datasets and (\textit{b}) USNA datasets. Solid black lines represent $\ds=\delta_{99}$. Dotted black lines represent ratios of $\ds/\delta_{99}$ from 1.1-1.4. $\ds$ was calculated by fitting to the Fourier model (filled symbols) and by linear interpolation (open symbols). Symbols for each dataset are given in table~\ref{tab:exps_published}.}
\label{fig7}
\end{figure}

We now demonstrate the applicability of our $\ds$ determination method on a range of experimental datasets, the parameters of which are given in table~\ref{tab:exps_published}. 
These datasets cover a range of Reynolds numbers as well as various non-canonical effects such as surface roughness, favourable-, and adverse-pressure gradients, of varying magnitudes and combinations. 
These datasets also cover two different single point measurement techniques, hot-wire anemometry and LDV, for comparison. 
The resulting skewness profiles, as a function of the wall-normal distance $z$ normalised by $\ds$, for selected representative cases, are shown in figure~\ref{fig6}. 
In each plot, the solid magenta lines represent the Fourier model (equation~\ref{eq:FM}) which was used to fit the data and determine $\ds$. 
Additionally, a simple linear interpolation between the two measurement points that bound the zero-crossing of the skewness profile, was also used to estimate $\ds$ and are compared with the results from using the Fourier model in figure~\ref{fig7}. 

Figure~\ref{fig6}(\textit{a}) shows hot-wire profiles of smooth-wall ZPG TBLs from MELB2. 
Darker colours indicate increasing streamwise measurement location and Reynolds number. 
The Fourier model appears to fit these skewness profiles well in the outer region of interest. 
In each case there are multiple data points between the negative and positive peaks in the skewness profile which can be used to fit the data with the Fourier model, even with conventional log-spaced measurement resolution (which is typically considered poor in the far outer region). 
Visually, the Fourier model also aligns well with a linear interpolation between the two points which bound the zero-crossing in the skewness profile. 
In each case, the linear interpolation gives a slightly higher estimate of $\ds$, owing to the shape of the skewness profile, but the relative difference in $\ds$ when using linear interpolation compared to fitting to the Fourier model is only $2.4\%$. 
This relative difference is the average difference between $\ds$ from the linear interpolation and the Fourier model, normalised by the $\ds$ from the Fourier model, for the six cases shown in each respective figure. 
This suggests that even with conventional experimental wall-normal spatial resolutions, and a somewhat random distribution of wall-normal measurement locations relative to the zero-crossing location, linear interpolation is still a relatively accurate alternative method to find $\ds$. 
Additionally, the difference between methods here is only an artefact of spatial resolution, and can be overcome in the future by doing well-resolved experiments. 

Figure~\ref{fig6}(\textit{b}) shows LDV profiles of smooth-wall pressure gradient TBLs from USNA1 \citep[Case 1: Stations 1, 2, 3, 9, 11 and 12 from][]{Volino_SmoothPG_2020}.
\rev{Shades of blue and red represent favourable- and adverse-pressure gradients, respectively, while shades of black represent ZPG conditions.} 
Darker colours indicate increasing streamwise measurement location and Reynolds number. 
For the LDV experiments, there is a significant level of Gaussian noise in the measurements \citep{Volino_SmoothPG_2020} which brings the magnitudes of the positive and negative peaks in the skewness back towards zero, as compared to the hot-wire profiles (figure~\ref{fig6}\textit{a}), or the Fourier model. 
However, these profiles maintain the shape of the typical skewness profile in the outer region, such that the Fourier model can still be used with some success. 
In the case of these measurements, linear interpolation appears to be more accurate by inspection, and is easier to implement compared to fitting with the Fourier model. 
The relative difference between the two methods for estimating $\ds$ is $10.2\%$ in this case, with the linear interpolation giving larger values of $\ds$. 
The consequence of this difference can be seen in figure~\ref{fig7}(\textit{b}) where comparison between $\ds$ and $\delta_{99}$ changes significantly depending on which method was used to determine $\ds$. 
The wall-normal spatial resolution of these measurements is typical, and similar to the hot-wire measurements. 
This is the highest relative difference observed between the two methods, for the present compilation of datasets, however this difference is less than the differences observed across the range of TBL thickness definitions currently used across the literature. 

Figure~\ref{fig6}(\textit{c}) shows hot-wire profiles of smooth-wall APG TBLs from MELB2. 
Darker colours indicate increasing streamwise measurement location, $\beta$, and Reynolds number. 
Similar to the ZPG cases above, the Fourier model appears to fit these profiles well, even with the addition of a non-canonical effect, namely an APG. 
Again, there is a relatively small difference of $4.7\%$ between using linear interpolation or the Fourier model for estimating $\ds$. 
These first three plots then demonstrate that the current methods are effective for smooth-wall pressure gradient TBLs, and are acceptable for use with various conventional single-point measurement techniques. 

Figure~\ref{fig6}(\textit{d}) shows hot-wire profiles of rough-wall ZPG TBLs from MELB4 for a single streamwise measurement location \citep[Case 1: $x=15$~m from][]{Squire_SmoothRoughHW_2016}. 
Darker colours indicate increasing $\ks$ and Reynolds number. 
Similar to the other hot-wire cases above, the Fourier model appears to fit well with data points which fall in the far outer region. 
Here, the relative difference between using the Fourier model or linear interpolation for estimating $\ds$ is also $4.7\%$. 
Importantly, this plot demonstrates that the profile of skewness in the outer region, for rough-wall TBLs, still has the critical feature (a zero-crossing) which allows us to use these methods to find $\ds$. 
Further, we expect that this will remain valid as long as the roughness sublayer does not reach the outer region of the TBL. 

Figure~\ref{fig6}(\textit{e}) shows LDV profiles of rough-wall pressure gradient TBLs from USNA3 \citep[Case 1: Stations 3, 4, 5, 9, 11 and 12 from][]{Volino_SmoothRoughPG_2023}. 
\rev{Shades of magenta and yellow represent favourable- and adverse-pressure gradients respectively, while shades of green represent ZPG conditions}. 
Darker colours indicate increasing streamwise measurement location, $\ks$, $\beta$, and Reynolds numbers. 
These profiles show the same behaviour as the previous LDV measurements where the negative and positive peak magnitudes are suppressed by measurement noise. 
However, the Fourier model fitting method still appears effective for finding $\ds$. 
Here, the relative difference between using the Fourier model or linear interpolation for estimating $\ds$ is only $4.5\%$. 
This case, and the ones before collectively, demonstrate that the proposed method can be applied to estimate the boundary layer thickness of both canonical and non-canonical TBLs measured using a variety of conventional techniques with typical experimental wall-normal resolutions. 

The values of $\ds$ which were found from figure~\ref{fig6}, combined with a selection of additional datasets from table~\ref{tab:exps_published}, are plotted against $\delta_{99}$ in figure~\ref{fig7} for reference. 
In figure~\ref{fig7}(\textit{a}), all hot-wire measurements conducted in the large Melbourne wind tunnel (MELB1-4) are compared. 
For the ZPG cases, both smooth- and rough-wall, the measured values of $\ds$ appear to fall between $1.2-1.3\delta_{99}$, consistent with the findings of \citet{Baxerres_PGBoundaryLayers_2024} (\ie $\Delta_{1.25}$). 
The main focus of this analysis is on experimental measurements, nonetheless the same comparison was also made for the ZPG simulation datasets (see table~\ref{tab:exps_published}). 
The same trend was observed, however these results are not plotted here due to large differences in the magnitude of $\delta$ (estimated in computational units) as compared with those in physical units plotted in figure~\ref{fig7}. 
PIV data points are included as cross symbols, and a good agreement with the hot-wire measurements is observed. 
In the case of the APG data, there is a deviation from the ZPG trend as the APG strength increases, as quantified by $\beta$ (which also corresponds with increasing TBL thickness). 
This emphasises the importance of finding $\ds$ directly from the skewness profile rather than employing an approximation based on $\delta_{99}$, for example, especially in pressure gradient TBLs. 
Additionally, there is good agreement in these trends regardless of the method used to estimate $\ds$ as shown by the open symbols (representing linear interpolation) and the filled symbols (representing the Fourier model fitting) in figure~\ref{fig7}(\textit{a}). 

\begin{figure}
\captionsetup{width=1.00\linewidth}
\begin{center}
\includegraphics[width=1.00\textwidth]{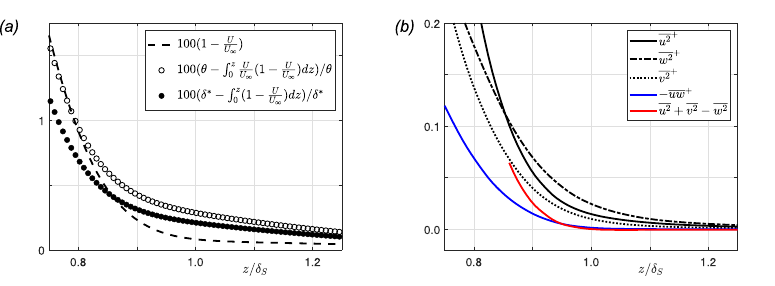}
\end{center}
\caption{\rex{(\textit{a}) Relative errors in computed momentum ($\theta$) and displacement ($\delta^{*}$) thicknesses as a function of the upper bound of integration and (\textit{b}) turbulent stress profiles near the TBL edge from KTH dataset \citep[LES of a ZPG TBL; ][]{EitelAmor_ZPGLES_2014}. Here $\delta_{99}=0.79\ds$.}}
\label{figB}
\end{figure}

In figure~\ref{fig7}(\textit{b}), a selection of LDV measurements conducted at the USNA are compared (multiple cases from USNA1-3 each). 
Because of differences in the magnitude of the boundary layer thickness and measurement techniques, these results have been plotted separately for comparison. 
In this case the trend between $\ds$ and $\delta_{99}$ changes depending on the method used to calculate $\ds$. 
It should be noted that the threshold used to find $\delta_{99}$ in these cases was modified (\ie $\scriptstyle \sqrt{\overline{u^{2}_{\infty}}} \textstyle /U_{\infty}=0.03$ which aligns with $U=0.99U_{\infty}$ for these cases) to account for the higher measurement noise, but was applied consistently between all USNA cases. 
When $\ds$ is found using the Fourier model, it tends to be close to $\delta_{99}$, highlighting the under prediction observed in figures~\ref{fig6}(\textit{c,e}). 
However, when linear interpolation is considered, the trends are very similar to those seen in figure~\ref{fig7}(\textit{a}), although there is more uncertainty, especially for the cases with complex/combined non-canonical effects. 
Due to the noise in the measurements and the lower magnitudes of $\delta$, errors in estimating both $\ds$ and $\delta_{99}$ may contribute to the uncertainty in these trends. 
Additionally, the streamwise variations in pressure gradient for the cases in figure \ref{fig7}(\textit{b}) means upstream pressure gradient history effects could also be responsible for the deviations from the consistent trend noted in figure \ref{fig7}(\textit{a}) (corresponding to experiments with minimal upstream pressure gradient history effects; \citealp{Lozier_APG_2024}). 
\rex{This is, however, a topic of ongoing research which will be addressed further in future studies. 
Nevertheless, these deviations reaffirm the importance of finding $\ds$ directly from a well-resolved measured skewness profile rather than relying on a predetermined relationship with another length scale, like $\delta_{99}$. 

These differences between $\ds$ and other TBL thickness definitions, like $\delta_{99}$, are also important to note as they can affect our characterisation of the TBL state, \egg through the calculation of integral length scales (\ie the displacement and momentum thicknesses). 
For example, by using $\delta_{S}$, instead of $\delta_{99}$, as the upper bound of integration for calculating the displacement and/or momentum thickness, the relative error in the calculation of these integrals reduces from approximately $1\%$ (when using $\delta_{99}$) down to less than $0.25\%$ (when using $\delta_{S}$) as shown in figure~\ref{figB}(a). 
This was true when tested on both numerical and experimental data, although this observation is more pertinent for conventional experimental measurements which are prone to errors in the calculation of these integral thicknesses.}

\rex{The suitability of $\ds$ for demarcating the TBL edge, compared to other common definitions such as $\delta_{99}$, can also be evaluated using theoretical predictions for the behaviour of turbulence statistics around the TBL edge. 
First, as mentioned previously, the wall-normal gradient of the streamwise velocity ($dU/dz$) is expected to go to zero at, and above, the TBL edge (for well-behaved canonical TBLs \cite{Coles_WakeParam_1956, Chauhan_CriteriaZPG_2009}). 
In figure~\ref{figB}(a), the mean velocity profile from the well-resolved KTH LES dataset \citep[dashed line;][]{EitelAmor_ZPGLES_2014} shows good agreement with this expected behaviour as the mean velocity reaches $> 99.8\%$ of $U_{\infty}$ at $z=\ds$ and the velocity gradient for $z>\ds$ is negligible. 
Additionally, \citet{Phillips_Irrotational_1955} predicted that the turbulent shear stresses should vanish at the boundary between a turbulent and irrotational flow (\ie at the TBL edge), while the normal stresses approach zero more gradually. 
From figure~\ref{figB}(b), the shear stress ($\overline{uw}^{+}$) does reach approximately zero at $z=\ds$ while the normal stresses continue to decay. 
\citet{Phillips_Irrotational_1955} also predicted that the mean energy of fluctuations normal to the boundary should be equal to the mean total energy of fluctuations parallel to the boundary (\ie at the TBL edge $\overline{w^2}^{+} = \overline{u^2}^{+} + \overline{v^2}^{+}$). 
Figure~\ref{figB}(b) confirms that this relationship between the normal stresses holds too as $\overline{u^2}^{+} + \overline{v^2}^{+} - \overline{w^2}^{+}$ is approximately zero for $z \geq \ds$. 
These above observations collectively demonstrate both the suitability and utility of using $\ds$ to define the TBL thickness.}

\subsection{Limitations and recommendations}
\label{sec:limitations}

Despite the demonstrated robustness in the methodologies adopted to implement the new $\ds$ definition, there are several limitations which have been mentioned previously and they will be discussed in further detail here. 
Accordingly, minor modifications are also recommended to be implemented in future experimental/numerical studies to enable and enhance the accuracy of the estimation of $\ds$ following the methods introduced here. 

\begin{figure}
\captionsetup{width=1.00\linewidth}
\begin{center}
\includegraphics[width=1.00\textwidth]{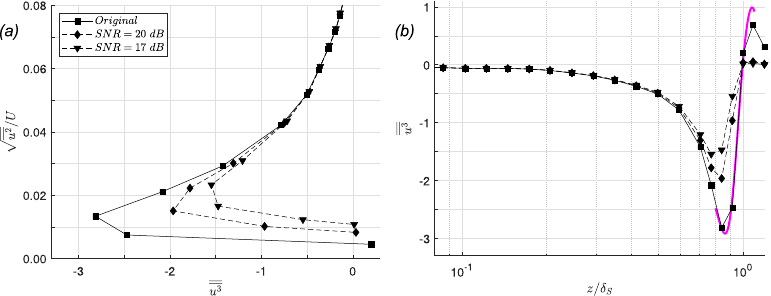}
\end{center}
\caption{(\textit{a}) Relationship between turbulence intensity and skewness and  (\textit{b}) wall-normal profiles of skewness from MELB2 with Gaussian white noise added to the experimentally measured time-series at varying signal-to-noise ratios.}
\label{fig8}
\end{figure}

The first limitation is related to high levels of measurement noise and/or freestream turbulence intensity. 
Figure~\ref{fig8} shows the effect of Gaussian white noise added to experimental velocity time series for a high-Reynolds number ZPG TBL (MELB2) at varying levels of signal-to-noise ratio (SNR). 
In figure~\ref{fig8}(\textit{a}) increasing SNR reduces the magnitude of the negative peak in the skewness, and also raises the apparent freestream turbulence intensity. 
In figure~\ref{fig8}(\textit{b}) it can also be seen that the artificial noise significantly reduces the magnitude of the positive peak in the skewness. 
However, due to the Gaussian behaviour of the artificial noise (\ie zero skewness is contributed), there is minimal effect observed on the location of the zero-crossing, and the same $\ds$ can be extracted from the profiles with added noise, as compared to the baseline.
The effect of noise can be further confirmed by considering the skewness profiles from the LDV measurements in figures~\ref{fig6}(\textit{b,e}), which have relatively high measurement noise (as compared to the baseline hot-wire measurements), and are comparable to the profiles in figure~\ref{fig8}(\textit{b}) with artificially added noise. 
Additionally, when freestream turbulence is sampled in the intermittent and freestream regions, an effect akin to adding Gaussian noise is expected on the skewness profile. 
This is highlighted in figure~\ref{fig8}(\textit{b}) where the compounding effects of the freestream turbulence (although small for the baseline hot-wire case) and artificial Gaussian noise lower the magnitude of the positive peak significantly. 
As such, it is recommended that both measurement noise and freestream turbulence levels be minimised so the crucial part of the skewness profile in the outer region can be resolved. 
It should be noted that at the other extreme, a sign change in the outer skewness profile is not observed in TBLs which have very high (induced) freestream turbulence intensities (ranging from 8-13\%) as shown in appendix~\ref{apx:B} \citep{Hearst_Freestream_2021}. 
At these levels of freestream turbulence intensity, it appears that the interface physics differs from the description given in $\S$~\ref{sec:definition} (and figure~\ref{fig2}), thereby making the application of our $\ds$ definition unsuitable for TBL flows with very high freestream turbulence intensity. 

\rev{A second limitation is related to errors in the measurement of skewness, \ie the convergence of higher-order statistics. 
For example, a relatively long sampling time is needed to achieve reasonable convergence of higher-order statistics, such as skewness, in conventional experimental measurements. 
This is especially true for measurement points around the skewness zero-crossing, as shown in figure~\ref{fig5}(\textit{c,e,f}), where infrequent, but significant, skewness-contributing events can alter the skewness magnitude when it is small. 
However, as long as this error is reasonable (\ie much less than the magnitude of the positive/negative peaks in the skewness), minimal adverse effects are expected on the estimation of $\ds$, as confirmed by figure~\ref{fig5}(\textit{c}). 
Conveniently, the statistical convergence of the skewness can be tested empirically, and improved through temporal, spatial and/or ensemble averaging of data. 
To that end, we recommend future measurements be designed with the convergence of higher-order statistics in mind to ensure a flow representative profile of the skewness.} 

\rev{A third limitation is related to the wall-normal spatial resolution of measurements, especially for conventional experimental measurement techniques. 
Moderate resolution, typical of conventional log-spaced experimental measurements, was found to be adequate for estimating the zero-crossing, as confirmed by figures~\ref{fig3} and \ref{fig6}. 
For reference, there were typically three measurement locations which fell between the negative and positive peaks in the skewness profile in these cases. 
However, with a higher spatial resolution (\ie more measurement locations in the outer region, see figure~\ref{fig5}), the skewness profile will be better resolved which is expected to lead to a more accurate estimation of the location of the zero-crossing. 
To quantify this, the skewness measurements of the ZPG TBL with high spatial resolution shown in figure~\ref{fig5} were considered as a baseline. 
Varying numbers of these measurement locations, which fall between the positive and negative peaks in the skewness profile, ranging from three up to the full available set, were then repeatedly selected at random and used to estimate $\ds$. 
For the minimum resolution, \ie using only three randomly distributed measurement locations between the positive and negative peaks, the simulated root mean square deviation of the measured $\ds$, from the baseline value of $\ds$, was found to be around 1\%. 
However, increasing the number of random locations considered, up to just five, nearly halved this simulated root mean square deviation. 
As such, we recommend making as many measurements as practical in the outer region (with particular focus on the region above $\delta_{99}$), ideally with 3-5 measurements at wall-normal locations falling between the negative and positive peaks in the skewness profile, to ensure high confidence when estimating the zero-crossing in the skewness profile using the proposed fitting method.} 

\rev{A final consideration, for future/previous numerical studies and low-order models in particular, is the requirement to resolve large-scale turbulent velocity fluctuations to some extent in order to compute the skewness. 
For example, it is not possible to extract skewness information from Reynolds Averaged Navier Stokes (RANS) simulations without the incorporation of an additional model for the skewness specifically, or a supplementary eddy-resolved simulation. 
However, in section~\ref{sec:AEM}, the results from an attached eddy model \citep{Deshpande_AEM_2021} are presented to demonstrate that even a relatively simple model, which statistically models the large, inertia-dominated eddies, without consideration of the non-linear interactions, can still resolve the outer skewness profile for the purposes of determining $\ds$.} 
\rew{There are, however, considerable differences observed at wall-normal locations near the wall between the skewness profiles from the AEM (figure~\ref{figA1}\textit{a}) and experimental measurements (figure~\ref{fig3}\textit{c}), which can be attributed to the disconsideration of small (viscosity-dominated) eddies and their non-linear interactions with the attached eddies in the AEM. 
It is also noted that consideration of additional TBL dynamics, not included in the AEM, such as intermittent small-scale turbulence, freestream turbulence, and TNTI dynamics, will further contribute to the skewness profile in the outer region. 
However, these simulations confirm the significance of large-scale motions in generating the unique features of the skewness profile in the TBL outer region, and provide guidance for the modelling required to reasonably resolve the outer skewness profile. 
Further investigation of the relationship between $\ds$, and other definitions for the TBL thickness (such as $\delta_{99}$), could also be useful for cases (such as RANS or laminar boundary layers) where some mean turbulence statistics are known, but the skewness is not.} 

Fortunately, it should be noted that many previously acquired datasets are already sufficient, with respect to the above limitations, to be reprocessed using this new definition/method, for comparison sake. 
However, the recommendations given here can also be used to design future experimental/numerical studies with the intention of applying this method to determine an appropriate and representative TBL thickness. 
Additionally, two different methods for estimating $\ds$ have been compared here, both of which were found to give reasonable results under the right conditions. 
As such, it is left to each analyst to choose an appropriate method of locating the skewness zero-crossing, based on the unique characteristics of their datasets/statistics. 
Finally, it is noted that the consideration of other turbulent flows of interest, such as jets or wakes, is beyond the current scope of this work. 
It is presently unclear if other turbulent shear flows will have similar and/or distinct skewness profiles with features which can be used to define a characteristic length scale in a way similar to the TBL. 
To this end, the results presented here can be used as motivation to revisit the definitions of characteristic length scales in other turbulent shear flows. 


\section{Summary}
\label{sec:summary}

A new statistical definition for the mean TBL thickness has been presented. 
By this definition, the TBL thickness is taken as the wall-normal location of the sign-change (or zero-crossing) in the streamwise velocity skewness profile (within the outermost region of the TBL). 
\rew{This new definition is motivated by the phenomenology of streamwise velocity fluctuations near the turbulent/non-turbulent interface, observed both experimentally and through attached eddy modelling, whose characteristics give rise to the distinct profile of skewness in the outer region of TBLs.} 
Furthermore, these characteristics are universal for any TBL that is developing under low freestream turbulence conditions (\ie irrespective of pressure gradients and/or surface roughness). 
This new definition is directly compared with previous definitions of TBL thickness, prevalent in the literature, using a recent large-scale experimental dataset which is uniquely suited to analysing the outer region of the TBL. 
The new definition not only yields a TBL thickness consistent with past definitions (\egg those based on Reynolds shear stress or `composite' mean velocity profiles), but it is also independent of any thresholds, by definition, and has been shown to be applicable to a range of conventional single-point measurements. 
In this way, the new definition can also be applied retroactively to the large body of TBL datasets that already exist in the literature. 
\rew{The new definition also yields a TBL thickness which can be used practically, for instance, to define an appropriate upper bound of integration when calculating integral thicknesses.} 
Additionally, two methods have been proposed to estimate the TBL thickness using this new definition: one based on linear interpolation of the measured skewness profile, and another based on fitting the measured skewness profile to a representative Fourier model of the general canonical skewness profile. 
The robustness, as well as the limitations, of these methodologies are demonstrated by employing various published experimental and numerical datasets, covering a broad range of canonical and non-canonical turbulent boundary layers, with varying degrees of wall-normal resolution and measurement noise. 
The relative difference between these methods is found to be less than the difference between the range of other prevalent definitions, suggesting either method can be used effectively for a variety of measurements. 
Several recommendations for future experiments and simulations are also given, namely higher spatial resolutions and better statistical convergence within the outer region to ensure this method can be applied successfully. 


\backsection[Acknowledgements]{
R.D.\ is grateful for financial support from the University of Melbourne's Postdoctoral Fellowship. 
\rev{The PhD research projects of co-authors L.L.\ and A.Z.\ were supported by an Australian Research Council Discovery Grant (DP210102172) and an Office of Naval Research (ONR) and ONR Global Grant (N62909-23-1-2068), respectively, under the supervision of co-authors I.M.\ and R.D. 
The research presented here is a practical extension, beyond the original scope of the aforementioned PhD research projects, which was primarily undertaken by M.L.\ with support from co-authors R.D. and W.A.R.} 
We are grateful to Profs.\ R. Volino, J. Jimenez and R.\ Vinuesa for sharing statistics from their published datasets, and we are grateful to Prof. H. M.\ Nagib for his comments on our preliminary draft. 
\rev{I.M.\ is also particularly grateful to Prof.\ N.\ Hutchins for insightful discussions on this topic over the past decade.}}


\backsection[Declaration of Interests]{The authors report no conflict of interest.}


\rev{\backsection[Author Contributions]{ \\
\textbf{Mitchell Lozier:} Methodology, investigation, formal analysis, validation, writing $-$ original draft.\\ 
\textbf{Rahul Deshpande:} Conceptualisation, methodology, writing $-$ review \& editing, funding acquisition, supervision.\\ 
\textbf{Ahmad Zarei:} Methodology, investigation, formal analysis\\ 
\textbf{Luka Lindi\'c:} Methodology, investigation, formal analysis\\ 
\textbf{Wagih Abu Rowin:} Methodology, formal analysis, writing $-$ review \& editing.\\ 
\textbf{Ivan Marusic:} Conceptualisation, methodology, writing $-$ review \& editing, funding acquisition, supervision.}}


\appendix

\section{Compatibility with high freestream turbulence levels}
\label{apx:B}

\begin{figure}
\captionsetup{width=1.00\linewidth}
\begin{center}
\includegraphics[width=1.00\textwidth]{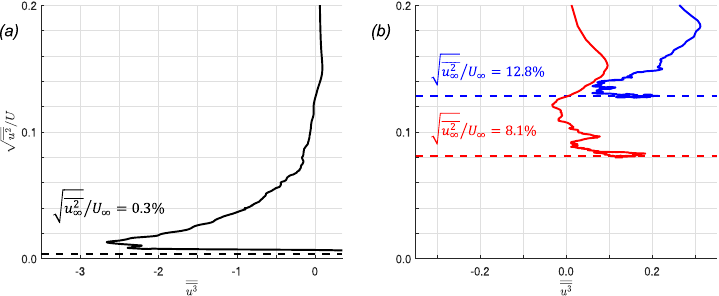}
\end{center}
\caption{Relationship between turbulence intensity and skewness for PIV measurements from (\textit{a}) a canonical ZPG TBL and (\textit{b}) TBLs with high levels of freestream turbulence \citep{Hearst_Freestream_2021}. \rex{Horizontal dashed lines represent the reported freestream turbulence intensity for each case.}}
\label{figA2}
\end{figure}

As mentioned in $\S$~\ref{sec:limitations}, there are limitations to the current method regarding high levels of freestream turbulence. 
In the recent large-scale experiments, described in $\S$~\ref{sec:exps_melb}, the freestream turbulence level was approximately $0.3\%$ ($\scriptstyle \sqrt{\overline{u^{2}_{\infty}}} \textstyle /U_{\infty}=0.003$) for the canonical case, shown by the dashed line in figure~\ref{figA2}(\textit{a}). 
This is typical of well-conditioned experimental facilities, however there are cases where the freestream turbulence may be (significantly) higher, based on the facility, or a result of being purposefully generated. 
Figure~\ref{figA2}(\textit{a}) shows the typical skewness profile on which we have based the current definition of TBL thickness. 
Skewness profiles from \citet{Hearst_Freestream_2021}, where the freestream turbulence level was purposefully made high ($8.1-12.8\%$), were also examined. 
In figure~\ref{figA2}(\textit{b}) we do not observe the characteristic skewness profile which is seen with minimal freestream turbulence (\ie figure~\ref{figA2}a), and in fact there is no longer a distinct zero-crossing in the outer region upon which to apply the new definition of boundary layer thickness. 
A comparison of figures~\ref{figA2}(\textit{a,b}) shows that in the case of very high freestream turbulence levels, the physical mechanisms of the outer region (\ie the TNTI) may have changed, and the phenomenological description given in $\S$~\ref{sec:definition} is no longer applicable. 
As such, in the case of these flows, an alternative definition of the boundary layer thickness should be employed. 

\bibliographystyle{jfm}
\bibliography{references}






\end{document}